\documentclass[galaxies,article,accept,moreauthors,10pt,a4paper]{mdpi}
\firstpage{1}
\articlenumber{x}
\doinum{10.3390/------}
\pubvolume{4}
\pubyear{2016}
\copyrightyear{2016}
\externaleditor{Academic Editors: Jose L. G\`{o}mez, Alan P. Marscher and Svetlana G. Jorstad}
\history{Received: 15 July 2016; Accepted: 18 August 2016 ; Published: date}

\usepackage{booktabs}
\usepackage{multirow}
\usepackage{soul}
\usepackage{microtype}
\newcommand\myurl[1]{\changeurlcolor{black}\url{#1}\changeurlcolor{blue}}
\makeatletter
\g@addto@macro{\UrlBreaks}{\UrlOrds}
\makeatother

 \theoremstyle{mdpi}
 \newcounter{thm}
 \newcounter{ex}
 \newcounter{re}
 \theoremstyle{mdpidefinition}

\Title{A Panchromatic View of Relativistic Jets in Narrow-Line Seyfert 1 Galaxies}

\Author{Filippo D'Ammando $^{1,2,}$*, Monica Orienti $^{2}$, Justin Finke $^{3}$, Josefin Larsson $^{4}$, Marcello Giroletti $^{2}$ and Claudia M. Raiteri $^{5}$}
\AuthorNames{Filippo D'Ammando, Monica Orienti, Justin Finke, Josefin Larsson, Marcello Giroletti  and Claudia M. Raiteri}

\address{%
$^{1}$ \quad {Dipartimento di Fisica e Astronomia}, Universit\'a degli Studi di Bologna, Viale Berti Pichat 6/2, \linebreak Bologna 40127, Italy \\
$^{2}$ \quad INAF - Istituto di Radioastronomia, Via Gobetti 101, Bologna 40129, Italy; orienti@ira.inaf.it (M.O.); \linebreak giroletti@ira.inaf.it (M.G.) \\
$^{3}$ \quad U.S. Naval Research Laboratory, Code 7653, 4555 Overlook Ave. SW, Washington, DC 20375-5352, USA; \linebreak justin.finke@nrl.navy.mil (J.F.) \\
$^{4}$ \quad KTH, Department of Physics, and the Oskar Klein Centre, AlbaNova, SE-106 91 Stockholm, Sweden; \linebreak josla@kth.se (J.L.) \\
$^{5}$ \quad INAF - Osservatorio Astrofisico di Torino, Via Osservatorio 20, I-10025 Pino Torinese (TO), Italy; \linebreak raiteri@oato.inaf.it (C.M.R.)}

\corres{Correspondence: dammando@ira.inaf.it}

\abstract{The discovery by the Large Area Telescope on board {\em Fermi} of variable $\gamma$-ray emission from radio-loud narrow-line Seyfert 1 (NLSy1) galaxies revealed the presence of a possible third class of Active Galactic Nuclei (AGN) with relativistic jets in addition to blazars and radio galaxies. Considering that NLSy1 are usually hosted in spiral galaxies, this finding poses intriguing questions about the nature of these objects and the formation of relativistic jets. We report on a systematic investigation of the $\gamma$-ray properties of a sample of radio-loud NLSy1, including the detection of new objects, using 7 years of {\em Fermi}-LAT data with the new Pass 8 event-level analysis. In addition we discuss the radio-to-very-high-energy properties of the $\gamma$-ray emitting NLSy1, their host galaxy, and black hole mass in the context of the blazar scenario and the unification of relativistic jets at different~scales.}


\conferencetitle{Blazars through Sharp Multi-Wavelength Eyes}

\begin{document}



\section{Introduction}

Since its launch on 11 June 2008, the {\em Fermi Gamma-ray Space Telescope}
has opened a new era in high-energy astrophysics. The primary instrument on
board {\em Fermi}, the Large Area Telescope~(LAT), is a~pair-conversion
telescope covering the energy range from $\sim$20 MeV up to $\sim$1 TeV with
unprecedented sensitivity and effective area~\cite{atwood09}. One of the major
scientific goals of the {\it Fermi} mission is to investigate the high-energy
emission in Active Galactic Nuclei (AGN) in order to understand the mechanisms
by which the particles are accelerated and the precise site of the
$\gamma$-ray emission. The combination of deep and fairly uniform exposure
over two orbits ($\sim$90 min), very good angular resolution, and stable
response of the LAT has allowed the production of the most sensitive, best-resolved survey of the $\gamma$-ray sky.

Before the launch of the {\em Fermi} satellite only two classes of AGN were
known to generate strong relativistic jets, and therefore to emit up to the
$\gamma$-ray energy range: blazars and radio galaxies, both hosted in giant
elliptical galaxies \cite{blandford78}. The first 4 years of observation by
{\em Fermi}-LAT confirmed that the extragalactic $\gamma$-ray sky is dominated
by blazars, with only a few radio galaxies detected \cite{acero15}. The~discovery by {\em Fermi}-LAT of variable $\gamma$-ray emission from a few
radio-loud narrow-line Seyfert 1 (NLSy1) galaxies revealed the presence of a possible third class of AGN with relativistic jets \cite{abdo2009a,abdo2009b}.

NLSy1 are a class of AGN identified by \cite{osterbrock85} and characterized
by their optical properties: narrow permitted emission lines (FWHM
(H$\beta$) $<$ 2000 km $\cdot$ s$^{-1}$), flux ratio [OIII]/H$\beta$ $<$ 3, and a bump due to Fe
II (e.g., \cite{pogge00}). They also exhibit strong X-ray variability, steep X-ray
spectra, substantial soft X-ray excess and relatively high luminosity
(e.g., \cite{grupe10}). These characteristics seem to point to systems with smaller
masses of the central black hole (BH; 10$^6$--10$^8$ M$_\odot$) and higher accretion rates
(close to or above the Eddington limit) with respect to blazars and radio
galaxies. NLSy1 are generally radio-quiet (radio-loudness $R$ $<$ 10), with only a small fraction ($<$ 7$\%$; \cite{komossa06}) classified as radio-loud. Objects with higher values of radio-loudness ($R$ $>$ 100) are even more sparse ($\sim$2.5\%), while $\sim$15$\%$
of quasars are~radio-loud. Considering that NLSy1 are usually hosted in spiral
galaxies, their detection in $\gamma$-rays poses intriguing questions about
the nature of these sources, the production of relativistic jets, and the
mechanisms of high-energy emission. In this paper we discuss the radio to
$\gamma$-ray properties of relativistic jets in NLSy1 galaxies.

In Section \ref{section 2}, we report the results of the LAT data analysis of a sample of NLSy1 over 7 years of {\em Fermi} observation, and discuss the $\gamma$-ray properties of NLSy1. In Sections \ref{section 3}-- \ref{section 5} we discuss the X-ray, infrared-to-UV, and radio properties, respectively, of the $\gamma$-ray NLSy1. Results about the modelling of the broad-band spectral energy distribution (SED) of $\gamma$-ray emitting NLSy1 are presentented in Section~\ref{section 6}. In Section \ref{section 7} we discuss about BH mass measurements, host galaxies, and the jet formation for these sources. Throughout the paper the photon indices are parameterized as $dN/dE \propto E^{-\Gamma_{\nu}}$, where $\Gamma_{\nu}$ is the photon index in the different energy bands. We adopt a $\Lambda$ cold dark matter cosmology with $H_0$ = 71 km $\cdot$  s$^{-1}$ $\cdot$ Mpc$^{-1}$, $\Omega_{\Lambda} = 0.73$, and $\Omega_{\rm m} = 0.27$ \cite{komatsu11}.

\section{The {\em Fermi}-LAT View of NLSy1} \label{section 2}

Four radio-loud NLSy1 galaxies have been detected at high significance by {\em Fermi}-LAT in the first year of operation ({i.e., 1H} 0323$+$342, PMN
J0948$+$0022, PKS 1502$+$036, and PKS 2004$-$447) and are included in the First LAT source catalog (1FGL) \cite{abdo10}. SBS 0846$+$513 was detected for the first time in $\gamma$-rays during October 2010--August 2011, when a significant increase in activity was observed by LAT~\cite{dammando12}. Thus in the Third LAT source catalog (3FGL) five NLSy1 are reported \cite{acero15}. The first $\gamma$-ray detection of FBQS J1644$+$2619 was
reported in \cite{dammando15}. An unidentified $\gamma$-ray source at 0.23 deg from the radio position of FBQS J1644$+$2619 was included in the
3FGL. Analysing the $\gamma$-ray data collected on a longer period than that of 3FGL, i.e., during August 2008--December 2014, the localization of the $\gamma$-ray source was better constrained. Moreover, the LAT detection of FBQS J1644$+$2619 in November 2008--January 2009  and in  July--October 2012 {corresponds} to periods of high optical activity, as observed in $V$-band by the Catalina Real-Time Sky Survey.

\subsection{{\em Fermi}-LAT Data Analysis}

In addition to these six NLSy1s already detected, new $\gamma$-ray emitting NLSy1 could be detected {after} accumulating more and more {\em Fermi}-LAT data. For this reason we analyze the first seven years of {\em Fermi}-LAT observations of a sample of 56 sources included in \cite{komossa06,yuan08,foschini11,foschini15} with a radio-loudness parameter higher than 10 in at least one of the previous samples. The LAT data used in this paper were collected from 5 August  2008  to 4 August 2015. During this time, the LAT instrument operated almost entirely in survey mode. The Pass 8 data \citep{atwood13}, based on a complete and improved revision of the entire LAT event-level analysis, were used. The analysis was performed with the
{ScienceTools} software package version v10r0p5. Only events belonging to the ``Source'' class ({evclass = 128}, \linebreak {\mbox{evtype = 3}}) were
used. We selected only events within a maximum zenith angle of 90 degrees to
reduce contamination from the Earth limb $\gamma$-rays. The spectral analysis
was performed with the instrument response functions {P8R2\_SOURCE\_V6} using
a binned maximum-likelihood method implemented in the Science tool
{gtlike}. Isotropic (``iso\_source\_v06.txt'') and Galactic diffuse emission
(''gll\_iem\_v06.fit'') components were used to model the background \citep{acero16}.
The normalization of both components was allowed to vary freely during the spectral fitting.

We analysed a region of interest of $30^{\circ}$ radius centred at the location of the NLSy1. We~evaluated the significance of the $\gamma$-ray
signal from the source by means of a maximum-likelihood test statistic~(TS) defined as TS = 2 $\times$ (log$L_1$ $-$ log$L_0$), where $L$ is the likelihood of the data given the model with ($L_1$) or without ($L_0$) a point source at the position of the target (e.g., \citep{mattox96}). The source model used in {gtlike} includes all the point sources from the 3FGL catalogue that fall within $40^{\circ}$ of \mbox{the target}. \mbox{The spectra} of these sources were parametrized by a power-law, a log-parabola, or a super exponential cut-off, as in the 3FGL catalogue. We also included new candidates within $10^{\circ}$ of the target from a preliminary source list using 7 years of Pass 8 data. A first maximum likelihood was performed to remove from the model the sources having TS $<$ 25. A second maximum likelihood was performed on the updated source model. As
a result, in addition to the six NLSy1 already detected by {\em Fermi}-LAT,
two new NLSy1 are detected with TS $>$ 25: B3 1441$+$476 (TS = 53), with a
photon index $\Gamma_\gamma$~=~2.65~$\pm$~0.15 and a 0.1--100 GeV flux of
(6.0 $\pm$ 1.9) $\times$ 10$^{-9}$ ph $\cdot$ cm$^{-2}$ $\cdot$ s$^{-1}$; NVSS J124634$+$023808
\linebreak (TS = 79),  $\Gamma_\gamma$ = 2.59 $\pm$ 0.09 and a 0.1--100 GeV flux of
(1.2 $\pm$ 0.2) $\times$ 10$^{-8}$ ph $\cdot$ cm$^{-2}$ $\cdot$ s$^{-1}$. No~significant detection was obtained for the other sources, with a 2-$\sigma$ upper limit ranging between \linebreak (0.6--7.8) $\times$10$^{-9}$ ph $\cdot$ cm$^{-2}$~$\cdot~$s$^{-1}$ (assuming~$\Gamma_\gamma$~=~2.5). More details will be reported in D'Ammando, Orienti et al. (in prep.).

\subsection{$\gamma$-ray Properties}

\begin{figure}[H]
\centering
\rotatebox{0}{\resizebox{!}{60mm}{\includegraphics{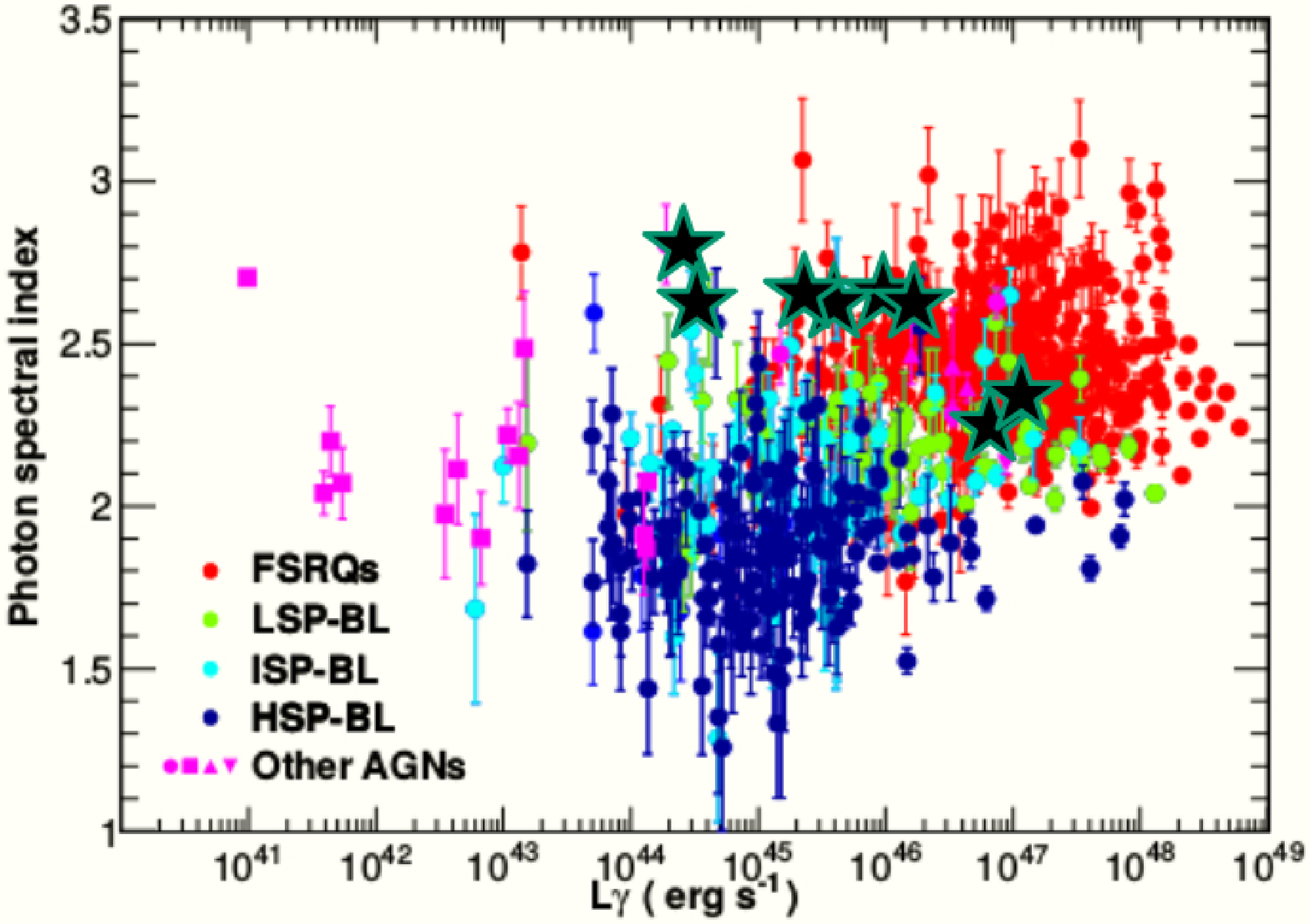}}}
\hspace{0.1cm}
\rotatebox{0}{\resizebox{!}{60mm}{\includegraphics{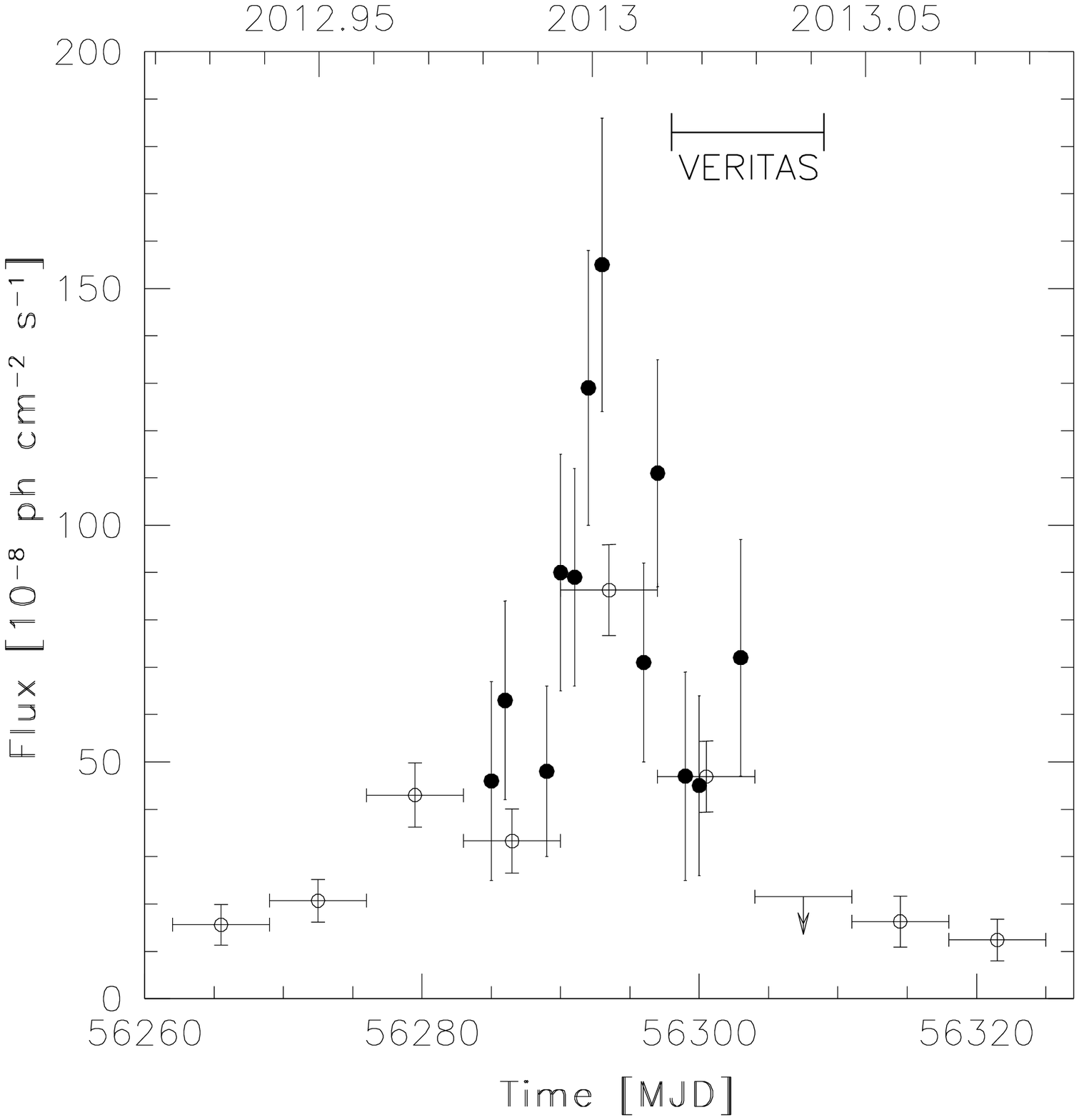}}}
\caption{{(\textbf{Left panel})}: Photon index versus $\gamma$-ray luminosity of
  different types of Active Galactic Nuclei (AGN) included in the~3LAC. Black stars represent the
  values obtained over 7 years for narrow-line Seyfert 1 (NLSy1). Adapted from~  \cite{ackermann15}; {(\textbf{Right~panel})}: LAT light curve of
  PMN J0948$+$0022 in the 0.1--100 GeV energy range during \mbox{1  December 2012}--31 January 2013
  with 7-day time bins. Arrow refers to 2-$\sigma$ { upper limit}. Upper
  limits are computed when $TS$ $<$ 10. { Red solid circles} represent daily fluxes reported for the periods of high activity. The horizontal line indicates the period of the VERITAS observation. Adapted from \cite{dammando15}.}
\label{LAT}
\end{figure}

The eight NLSy1 detected by {\em Fermi}-LAT at high significance up to now span a redshift range between 0.061 and 0.705. The average apparent $\gamma$-ray isotropic luminosity of these sources in the 0.1--100 GeV energy band is between 10$^{44}$ and 10$^{47}$ erg $\cdot$ s$^{-1}$, a range of values typical
of blazars. \mbox{This may be} an indication of a small viewing angle with respect to
the jet axis, and therefore a high beaming factor for the $\gamma$-ray
emission, similarly to blazars. In the same way, the average photon index
$\Gamma_\gamma$ ranges between 2.2 and 2.8, a range of values usually observed
for blazars \cite{ackermann15}. In Figure \ref{LAT} \mbox{(left panel)},
we compare the $\gamma$-ray photon index and luminosity obtained over 7
years for the eight NLSy1 with those obtained over 4 years for the AGN included in the Third LAT Catalog of AGN (3LAC)~\cite{ackermann15}. \mbox{It is clear that} the NLSy1 lie in the region occupied by
flat spectrum radio { quasars} (FSRQ), with the exception of 1H 0323$+$342 and FBQS J1644$+$2619. However, these two sources have the lowest redshift among the eight NLSy1, and this might be the reason why their $\gamma$-ray luminosity is lower than the other NLSy1 and FSRQ.

Moreover, several strong $\gamma$-ray flares were observed from SBS 0846$+$513, PMN J0948$+$0022, PKS 1502$+$036, and 1H 0323$+$342, reaching a
peak apparent isotropic $\gamma$-ray luminosity of \linebreak $\sim$10$^{48}$ erg $\cdot$ s$^{-1}$~\cite{foschini11b,dammando13,paliya14,dammando15b,dammando16},
comparable to that of bright FSRQ~\cite{ackermann15}. Variability was observed
during all $\gamma$-ray flares on a daily time-scale{; in case of} PKS 1502$+$036 sub-daily variability was observed~\cite{dammando16}. \mbox{Moreover, SBS 0846}$+$513 and PMN J0948$+$0022 showed a $\gamma$-ray flaring activity combined with a moderate spectral evolution, a behaviour that was already observed in bright FSRQ and low-synchrotron-peaked BL Lacs \cite{abdo10b}. To summarize, luminosity, variability and spectral properties of these NLSy1 in $\gamma$-rays indicate a blazar-like behaviour.

NLSy1 are included neither in the First Fermi LAT Catalog of sources above 10 GeV (1FHL; \cite{ackermann13}) nor in the Second Catalog of
hard Fermi-LAT Sources above 50 GeV (2FHL; \cite{ackermann16}). At Very High Energy (VHE; E $>$ 100 GeV), VERITAS observations of PMN
J0948$+$0022 were carried out a few days after the $\gamma$-ray flare observed by {\em Fermi}-LAT on 1 January  2013
(Figure \ref{LAT}, right panel). These observations resulted in an upper limit of F$_{> 0.2\rm\,TeV}$ $<$ 4 $\times$ 10$^{-12}$ ph $\cdot$ cm$^{-2}$ $\cdot$ s$^{-1}$ \cite{dammando15b}. The lack of detection at VHE could be due to different reasons: (1) The distance of the source ($z$ = 0.5846) is relatively large and most of the GeV/TeV emission may be absorbed due to pair production from $\gamma$-ray photons of the source and the infrared photons from the extragalactic background light;
(2) The VERITAS observations were carried out a few days after the peak of the
$\gamma$-ray activity, thus covering only the last part of the MeV/GeV~flare; (3) Considering the similarities with FSRQ, a broad line region (BLR)
should be present in these NLSy1. The presence of a BLR could produce a
spectral break due to pair production, suppressing the flux beyond a few GeV
and preventing a VHE detection. Future observations with the Cherenkov
Telescope Array (CTA) will constrain the level of $\gamma$-ray emission at 100 GeV or below.

\section{X-ray Properties}\label{section 3}

The X-ray spectra of NLSy1 are usually characterized by a soft photon index
($\Gamma_{\rm\,X}$ $>$ 2, \cite{grupe10}). \mbox{On the contrary}, a relatively hard
X-ray spectrum was detected in the {\em Swift}/XRT observations of SBS
0846$+$513~\cite{dammando12,dammando13}, PMN J0948$+$0022 \cite{foschini11b, dammando14, dammando15b}, 1H 0323$+$342 \cite{paliya14}, PKS
1502$+$036~\cite{dammando13b,dammando16}, and PKS 2004$-$447~\cite{orienti15,kreikenbohm16}. This suggests a significant contribution of
inverse Compton radiation from a relativistic jet in X-rays, similar to what
is found for FSRQ~(e.g., \cite{dammando11}).

The high quality of the {\em XMM-Newton} observation of PMN J0948$+$0022 performed in
\mbox{28--29 May 2011}  allowed us to study in detail its X-ray spectrum, as reported
in \cite{dammando14}. The spectral fitting of the {\em XMM-Newton} data shows
that emission from the jet most likely dominates the spectrum above $\sim$2
keV, while a soft X-ray excess is evident below $\sim$2 keV. The origin of the soft X-ray excess is
still an open issue \cite{gierlinski04}. This component is a typical feature in the
X-ray spectra of radio-quiet NLSy1, but it is quite unusual in jet-dominated
AGN, with some exceptions (e.g., PKS 1510$-$089; \cite{kataoka08}). In the case of
PMN J0948$+$0022, the statistics did not allow us to distinguish between
different models for the soft X-ray emission. Models where the soft emission
is partly produced by blurred reflection, or Comptonisation of the thermal
disc emission, or simply a steep power-law, all provide good fits to the
data. A multicolor thermal disc emission also gives a comparable fit, but the
temperature is too high (kT = 0.18 keV) and is incompatible with a standard
Shakura \& Sunyaev accretion disc \cite{dammando14}. A~clear soft X-ray
{emission component} is observed also in the {\em XMM-Newton} observation of 1H 0323$+$342
collected in 23--24 August 2015   (D'Ammando et al., in prep.).

An indication of a weak soft excess was observed in the {\em XMM-Newton}
spectrum of PKS 2004$-$447 in 2004~\cite{gallo06}. On the contrary, the X-ray
spectra of PKS 2004$-$447 collected in  May and October 2012 are well
reproduced by a single power-law with a hard photon index ($\Gamma_{\rm\,X}$ $\sim$1.7) and no significant soft X-ray excess~\cite{orienti15,kreikenbohm16}. The {\em XMM-Newton} spectrum of PKS 1502$+$036 collected in 2012 was quite well fit by a simple power-law model, although some residuals are observed at low and high energies. These residuals might be a hint of the presence of the soft X-ray excess and the Iron line, respectively. A better fit was obtained by using a broken power-law model, suggesting the presence of two emission components in X-rays, but the uncertainties related to the spectral parameters are quite large~\cite{dammando16}.

Only one source, 1H 0323$+$342, among the NLSy1 detected by {\em Fermi}-LAT
was detected in hard X-rays ($E>$ 10 keV). This source was included in the
70-month {\em Swift}-BAT catalogue, with a photon index of 1.73 $\pm$ 0.25
\cite{baumgartner13}. 1H 0323$+$342 was included also in the Fourth IBIS/ISGRI
Soft Gamma-ray Survey Catalog \cite{bird10}.

\section{Infrared-to-UV Properties} \label{section 4}

Thanks to the all sky survey carried out by the {\em Wide-field Infrared Survey (WISE)} it was possible to study from a statistical point of view the infrared properties of blazars, dominated by the jet non-thermal emission,
in comparison with the other extragalactic sources that are dominated by
thermal emission (e.g., \cite{dabrusco12}). A dedicated study of the infrared
colours of NLSy1 indicates that sources detected by LAT show infrared
properties similar to FSRQ, with a dominance of the jet emission component in
the infrared band \cite{caccianiga15}. By investigating {\em WISE} data,
\cite{jiang12} found infrared intraday variability in SBS 0846$+$513 and PMN
J0948$+$0022, with an amplitude of $\sim$0.1--0.2 mag. A similar variability
amplitude was observed for PKS 1502$+$036 on a monthly time-scale.

Optical intraday variability has been reported for PMN J0948$+$0022 \cite{maune13, itoh13}, SBS 0846$+$513 \cite{maune14}, and 1H
0323$+$342 \cite{paliya14}. In the case of PMN J0948$+$0022 and SBS 0846$+$513
the intraday variability is sometimes associated with a significant increase
of the optical polarisation percentage, indicating the relativistic jet as the
main origin for the optical emission in these objects. The optical and UV part of the spectrum of FSRQ is usually dominated by the
thermal emission of the accretion disc, while in BL Lac objects the disc is radiatively inefficient, with the disc emission overwhelmed by the non-thermal jet emission (e.g., \cite{ghisellini11}). The accretion disc emission is visible in the low activity state of the SED of PMN J0948$+$0022 \cite{dammando15b}, 1H 0323$+$342 \cite{abdo2009b} and PKS 1502$+$036 \cite{dammando16}.  On the contrary, no significant evidence of thermal emission from the accretion disc has been observed in SBS 0846$+$513~\cite{dammando13} and PKS 2004$-$447~\cite{orienti15}.

\section{Radio Properties} \label{section 5}

On pc scale a core-jet structure was observed for SBS 0846$+$513 \cite{dammando12}, PKS 2004$-$447~\cite{orienti15}, 1H~0323$+$342~\cite{wajima14}, PKS 1502$+$036~\cite{dammando13b}, and PMN J0948$+$0022~\cite{giroletti11, dammando14}. The analysis of the 6-epoch data set of SBS 0846$+$513 collected by the MOJAVE programme during 2011--2013
indicates that a superluminal jet component is moving away from the core with
an apparent angular velocity of (0.27~$\pm$~0.02) mas $\cdot$  year$^{-1}$, corresponding
to (9.3 $\pm$ 0.6)$c$ \cite{dammando13}. This apparent superluminal velocity
indicates the presence of boosting effects for the jet of SBS
0846$+$513. Apparent superluminal velocity of a jet component was reported also for PMN J0948$+$0022 and 1H 0323$+$342 \cite{lister16}.
\mbox{On the contrary}, VLBA observations did not detect apparent superluminal motion at 15 GHz for PKS 1502$+$036 during 2002--2012, although the radio spectral variability, the one-sided jet-like structure, the observed $\gamma$-ray luminosity and the Doppler factor estimated by SED modelling seem to require the presence of boosting effects in a relativistic jet \cite{dammando13}. A sub-luminal component was reported also in~\cite{lister16}. \mbox{This result resemble}s the `Doppler factor crisis' observed in bright TeV BL Lacs (e.g., \cite{piner10}).

Strong radio variability was observed at 15 GHz during the monitoring of the OVRO 40-m telescope of PMN
J0948$+$0022~\cite{dammando14,dammando15b}, PKS 1502$+$036 \cite{dammando13b}, SBS
0846$+$513~\cite{dammando12,dammando13}, and 1H 0323$+$342~\cite{paliya14}. An inferred variability brightness
temperature of 2.5 $\times$ 10$^{13}$~K, 1.1 $\times$ 10$^{14}$~K, and 3.4~$\times$~10$^{11}$~K was obtained for PKS 1502$+$036, SBS 0846$+$513, and
PMN J0948$+$0022, respectively. These values are larger than the brightness temperature derived for the
Compton catastrophe~\cite{readhead94}, suggesting that the radio emission of the jet is Doppler boosted. On the
other hand, a high variability brightness temperature of 10$^{13}$ K, comparable to that of the $\gamma$-ray
NLSy1, was observed for TXS 1546$+$353. However, no $\gamma$-ray
emission has been detected from this source, so far
\cite{orienti15}. Moreover,~{ intensive} monitoring of the $\gamma$-ray NLSy1 from 2.6 GHz to 142 GHz
with the Effelsberg 100-m and IRAM 30-m telescopes showed, in addition to an
intensive variability, spectral evolution across the different bands
following evolutionary paths explained by travelling shocks, typical
characteristics seen in blazars \cite{angelakis15}.

On kpc scale a two-sided radio structure was detected for PMN J0948$+$0022, 1H
0323$+$342, and FBQS J1644$+$2619 \cite{doi12}, with a core-dominated
structure as observed in blazars.

\section{SED Modelling of $\gamma$-ray Emitting NLSy1} \label{section 6}

\begin{figure}[H]
\centering
\rotatebox{0}{\resizebox{!}{50mm}{\includegraphics{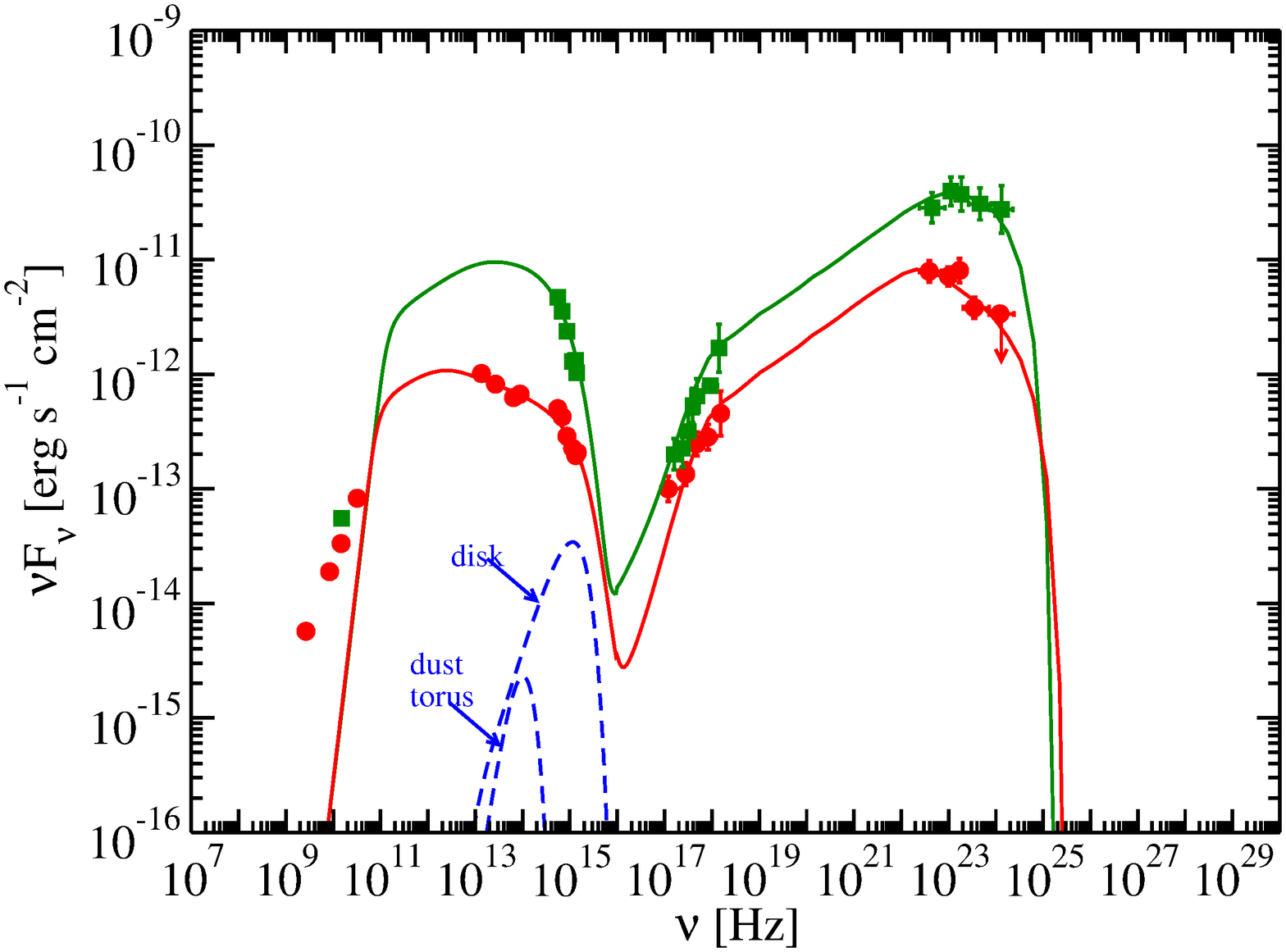}}}
\hspace{0.3cm}
\rotatebox{0}{\resizebox{!}{48mm}{\includegraphics{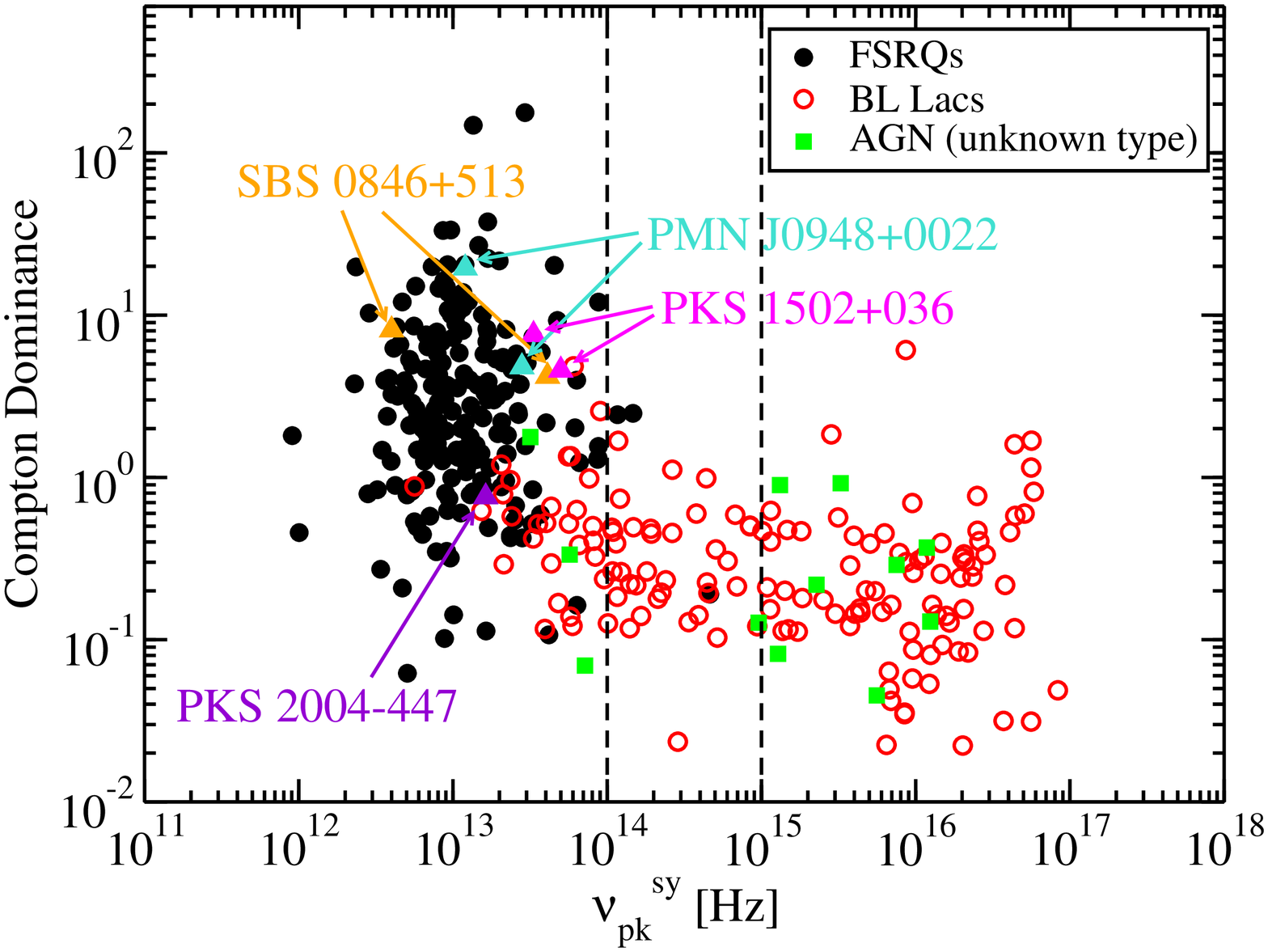}}}
\caption{{ (\textbf{Left panel})}: Spectral energy distribution (SED) and model fit (solid curve) of SBS\,0846$+$513 in flaring activity (squares) and low activity (circles) with the { disk and dust torus emission components} shown as dashed curves. Adapted from \cite{dammando13}; { (\textbf{Right panel})}: the Compton dominance versus peak synchrotron frequency for sources from the Second LAT AGN catalog (2LAC) \cite{ackermann11}, along with two activity states of SBS 0846$+$513~\cite{dammando13}, PMN J0948$+$0022 \cite{dammando15b}, PKS 1502$+$036 \cite{dammando16}, and an average state of PKS 2004$-$447
\cite{orienti15}. Dashed lines indicate the boundary between
high-synchrotron-peak, intermediate-synchrotron-peaked, and
low-synchrotron-peaked blazars. Adapted from \cite{finke13}, reproduced by permission of the AAS.}
\label{SED}
\end{figure}

The first SED collected for the NLSy1 detected in the first year of {\em Fermi} operation showed clear similarities with blazars: a double-humped
shape with a first peak in the IR/optical band due to synchrotron emission, a
second peak in the MeV/GeV band likely due to inverse Compton emission. The
physical parameters { of the jet of} these NLSy1 are blazar-like, and the jet power is in the average range of blazars \cite{abdo2009b}. We compared the SED of SBS 0846$+$513 during the flaring state in May 2012  with that of a quiescent state (Figure \ref{SED}, left panel). The SED of the two different activity states, modelled by an external Compton (EC) component of seed photons from a dust torus, could be fitted by changing the electron distribution parameters as well as the magnetic field \cite{dammando13}. A significant shift of the synchrotron peak to higher frequencies was observed during the May 2012  flaring episode, similar to FSRQ (e.g., PKS 1510$-$089; \cite{dammando11}).

For PMN J0948$+$0022 we compared the broad-band SED of the 2013 flaring activity state with that from an intermediate activity state observed in
2011. Contrary to what was observed for some FSRQ (e.g., PKS 0537$-$441; \cite{dammando13c}) the SED of the two activity states, modelled as
synchrotron emission and EC scattering of seed photons from a dust torus, could not be modelled by changing only the electron
distribution parameters. A higher magnetic field is needed for the high
activity state, consistent with the modelling of different activity states of
PKS 0208$-$512 \cite{chatterjee13}. We also modelled the 2013 flaring state of
PMN J0948$+$0022  assuming Compton-scattering of BLR line radiation. The model
reproduces the data as well as the scattering of the IR torus photons, but
requires magnetic fields which are {far from equipartition between the energy densities of particles and magnetic field}.
In the case of PKS 2004$-$447, the inverse Compton component is modelled with an EC scattering of dust torus seed
photons, as for the SED of SBS 0846$+$513 and PMN J0948$+$0022 \cite{orienti15}.
For PKS 1502$+$036 the SED of an average activity state was compared with that
of the 2015 flaring state. The two SED could be fitted with the high-energy
bump modelled as an EC component with seed photons from a dust
torus, and by changing the electron distribution parameters as well as the
magnetic field. The fit of the disc emission during the average state
constrains the BH mass to values lower than 10$^8$ M$_{\odot}$ \cite{dammando16}.

Following the 'classical' blazar sequence \cite{ghisellini98}, \cite{finke13} found a clear correlation between Compton dominance and the
rest-frame peak synchrotron frequency ($\nu_{pk}^{sy}$) in blazars, related
to the contribution of the EC component to the high-energy
bump. In Figure \ref{SED} (right panel) the Compton dominance versus
$\nu_{pk}^{sy}$ obtained for the sources in the 2LAC is
compared to the values obtained for 5 $\gamma$-ray emitting NLSy1 in different
activity states. NLSy1 lie in the region occupied by FSRQ, confirming that the EC is the dominant mechanism for producing the $\gamma$-ray
emission in these sources, and therefore the similarity between FSRQ and these NLSy1.

\section{Black Hole Mass, Host Galaxy, and Jet Formation} \label{section 7}

The mechanisms for producing a relativistic jet are still unclear. In
particular, the physical parameters that drive the jet formation are still
under debate. One of the key parameters should be the BH mass, with
only large masses allowing an efficient jet formation
(e.g., \cite{sikora07}). The most powerful jets are found in luminous
elliptical galaxies with very massive BH. This was interpreted as an indirect
evidence that a high spin is required for the jet production, since at least
one major merger seems to be necessary to spin up the SMBH \cite{sikora07, chiaberge11} suggested that a BH mass higher than 10$^{8}$ M$_\odot$ is necessary for producing a radio-loud AGN and that the merger history together with the subsequent galaxy morphology plays a fundamental role.

The detection of variable $\gamma$-ray emission from a few NLSy1 galaxies confirmed the presence of a relativistic jet in these
objects. The discovery of relativistic jets in a class of AGN usually hosted in spiral galaxies (e.g., \cite{deo06}), with a BH mass
ranging between 10$^{6}$ and 10$^{8}$ M$_\odot$, was a great surprise. This~discovery challenges the current knowledge on how the
structures are generated and developed (e.g., \cite{bottcher02,
  marscher09}). However, {the estimated masses of the NLSy1 have} large uncertainties. It was suggested that the
BH masses of NLSy1 are underestimated due either to the effect of radiation
pressure \cite{marconi08} or to projection effects \cite{baldi16}. Higher BH masses than those derived by the virial method (e.g., \cite{yuan08})
 are in agreement with the values estimated by modelling the optical/UV data
 with a Shakura and Sunyaev disc spectrum (i.e., 10$^{8}$--$10^{9}$ M$_\odot$; \cite{calderone13}). This may solve the problem of the minimum
BH mass predicted in different scenarios of relativistic jet formation, but introduces a possible new issue. Spiral galaxies are formed by minor mergers, with BH mass typically ranging between 10$^6$ and 10$^8$ M$_\odot$ (e.g., \cite{woo02}). If~the BH mass for radio-loud NLSy1 is on the larger side of the estimated values, how is it possible to reconcile such a large BH mass with a spiral galaxy not subject to major mergers?

Among the radio-loud NLSy1 detected in $\gamma$-rays up to now, only for the closest one, {1H~0323}$+$342 ($z$ = 0.061), the host galaxy structure was detected. Observations with the {\em Hubble Space Telescope} and the {\em Nordic Optical Telescope} revealed a one-armed galaxy morphology or a
circumnuclear ring, respectively, suggesting two possibilities: the spiral arm of
the host galaxy~\cite{zhou07} or the residual of a galaxy merger~\cite{anton08, leon14}. These observations, together with the~lack of information about the host galaxy of the other $\gamma$-ray emitting NLSy1,
{ leave room} for the hypothesis that the NLSy1 detected in $\gamma$-rays could have
peculiar host galaxies with respect to the other NLSy1 (e.g., non-spiral morphology,
undergoing strong merger activity). Further high-resolution observations of the host
galaxy of $\gamma$-ray emitting NLSy1 are fundamental to obtain important insights into
the onset of production of relativistic jets in these sources.

\vspace{6pt}


\acknowledgments{The \textit{Fermi}-LAT Collaboration acknowledges support for LAT development, operation and data analysis from NASA and DOE (United States), CEA/Irfu and IN2P3/CNRS (France), ASI and INFN (Italy), MEXT, KEK, and JAXA (Japan), and the K.A.~Wallenberg Foundation, the Swedish Research Council and the National Space Board (Sweden). Science analysis support in the operations phase from INAF (Italy) and CNES (France) is also gratefully acknowledged.}


\authorcontributions{F.D. analyzed the X-ray and $\gamma$-ray data, and wrote the paper; M.O. analyzed the radio and $\gamma$-ray data, and contributed to write the paper; J.F. modelled the SED; J.L. analyzed the X-ray data; M.G. analyzed the radio data; C.M.R. analyzed the optical and UV data.}   
\conflictofinterests{The authors declare no conflict of interest.}




\bibliographystyle{mdpi}

\renewcommand\bibname{References}



\end{document}